\documentstyle[12pt,aaspp4]{article}
\lefthead{Cochran {\it et~al.\/}}
\righthead{Planetary Companion to 16 Cyg B}
\begin{document}
\slugcomment{WARNING: Unrefereed Version}

\title{The Discovery of a Planetary Companion to 16 Cygni B}

\author{William D. Cochran, Artie P. Hatzes}
\affil{The University of Texas at Austin}
\authoraddr{McDonald Observatory, Austin, TX, USA 78712}
\authoremail{wdc@astro.as.utexas.edu, artie@astro.as.utexas.edu}

\author{R. Paul Butler\altaffilmark{1}, Geoffrey W. Marcy\altaffilmark{1}}
\affil{San Francisco State University}
\authoraddr{Department of Physics and Astronomy, San Francisco, CA, USA 94132}
\authoremail{gmarcy@etoile.berkeley.edu, paul@further.berkeley.edu}
\altaffiltext{1}{Also at Department of Astronomy, University of California,
Berkeley, CA USA  94720}

\begin{abstract}

High precision radial velocity observations of the solar-type star
16~Cygni~B (HR\,7504, HD\,186427), taken at McDonald Observatory and
at Lick Observatory,
have each independently discovered periodic radial-velocity variations
indicating the presence of a Jovian-mass companion to this star.
The orbital fit to the combined McDonald and Lick data gives a period
of 800.8\,days, a velocity amplitude (K) of 43.9\,m\,s$^{-1}$, and
an eccentricity of 0.63.
This is the largest eccentricity of any planetary system discovered so far.
Assuming that 16~Cygni~B has a mass of 1.0M$_{\odot}$, the mass function
then implies a mass for the companion of $1.5/\sin i$ Jupiter masses.
While the mass of this object is well within the range expected for planets,
the large orbital eccentricity cannot be explained simply by the standard
model of growth of planets in a protostellar disk.
It is possible that this object was formed in the normal manner with a low
eccentricity orbit, and has undergone post-formational orbital evolution,
either through the same process which formed the ``massive eccentric'' planets
around 70~Virginis and HD114762, or by gravitational interactions with the
companion star 16~Cygni~A.
It is also possible that the object is an extremely low mass brown dwarf,
formed through fragmentation of the collapsing protostar.
We explore a possible connection between stellar photospheric Li depletion,
pre-main sequence stellar rotation, the presence of a massive proto-planetary
disk, and the formation of a planetary companion.
\end{abstract}


\section{Introduction}
\label{intro}
A decade-long effort to detect sub-stellar companions to solar-type stars
by radial-velocity techniques gave the first tantalizing hints
with the discovery of a low-mass object in orbit around HD\,114762 by Latham
{\it et~al.\/} (1989) (cf. \cite{CoHaHa91}; \cite{Wi96}). \nocite{LaMaSt89}
Several more years of intense effort by various groups (\cite{WaWaIr95};
\cite{MMMoPe94}; \cite{CoHa94}; \cite{MaBu92}; \cite{MaQuUd96})
have finally achieved a stunning success beginning with the discovery
of radial-velocity variations implying the presence of a planetary companion
to 51~Pegasi (\cite{MaQu95}; \cite{MaBuWi97}),
followed in short order by the detection of companions to
70~Virginis (\cite{MaBu96}),
47~Ursae~Majoris (\cite{BuMa96}),
$\rho^1$~Cancri, $\tau$~Bo\"otes, $\upsilon$~Andromedae (\cite{BuMaWi96b}),
and the tantalizing but unconfirmed astrometric companion to
Lalande 21185 (\cite{Ga96}).
The systems discovered so far can be categorized roughly into three different
classes: 1) the ``51-Peg'' type planets (the companions to 51~Peg, 55~Cnc,
$\tau$~Boo, and $\upsilon$~And) which have minimum masses
around one Jovian mass and orbital periods of several days, 2) the ``massive
eccentric'' objects around HD114762 and 70~Vir, with minimum masses of 6-10
M$_{J}$, semi-major axes of 0.4--0.5\,{\sc au}, and orbital eccentricities of
0.3--0.4, and 3) the ``pseudo-Jovian'' planets around 47~UMa and
Lalande 21185, with low eccentricity, masses up to a few Jovian masses,
and semi-major axes over 2\,{\sc au}.
None of these systems resemble our own solar system very closely; they all
have a Jovian planet much closer to the parent star.
However, there are strong observational selection effects which led to the
discovery of these types of system before Jovian planets in wider orbits
are detected.
Since the observed stellar radial-velocity signal is proportional to the
companion object mass, and is inversely proportional to the square root of the
semi-major axis, massive planets in close orbits will give the largest stellar
reflex velocities and thus will be the first systems detected.   The detection
of systems such as our own can be done with the precision of current surveys,
but will require a longer baseline of observations.

We report here the detection of a planetary-mass companion to the solar-type
star 16~Cygni~B, the secondary star of the 16~Cygni triple star system.
This object has a minimum mass well within the range of that expected on
theoretical grounds for ``planets'', a semi-major axis which places the object
near the ``habitable zone'' (\cite{KaWhRe93}), but with an extremely
large orbital eccentricity.
The low mass, coupled with the high eccentricity, makes this
planetary companion unlike any of the previously discovered systems.

\section{Observations}
\label{obs}
Following the inspiration of the pioneering precise radial-velocity program
of Campbell and Walker (\cite{CaWa79}; \cite{CaWaYa88}),
three different high-precision
radial velocity programs started major surveys for sub-stellar companions to nearby solar-type stars in 1987 (\cite{MMMoPe94}; \cite{CoHa94}; \cite{MaBu92}).
The Lick Observatory and McDonald Observatory programs both included the
star 16~Cygni~B (HR\,7504, SAO\,31899, HD\,186427, BD+50~2848) in their
surveys.
This is the secondary star in a system comprising a pair of G~dwarfs in a wide
visual binary, and a distant M dwarf.
Table~\ref{star_properties} compares both 16~Cygni~A and 16~Cygni~B to the Sun.
Both of the G~dwarf stars in the 16~Cygni system have effective temperatures,
masses, surface gravities, and heavy element abundances very close
to the solar value.
This clearly demonstrates why both stars are widely regarded as excellent
``solar-analog'' stars (\cite{Ha78}; \cite{CSKnHe81}; \cite{FrCSCh93};
\cite{Gr95}).
The spectrum of 16~Cygni~B is almost identical to that of the Sun, making
this star virtually a solar twin.


The observational techniques used in the Lick and McDonald surveys to
achieve extremely high radial-velocity precision are
discussed in detail by Marcy and Butler (1992) and by Cochran and Hatzes
(1994) respectively.   The McDonald survey started in 1987 using the
telluric O$_2$ at 6300{\AA} as a high-precision radial-velocity metric
(\cite{GrGr73}).
The survey switched to the use of an I$_2$ gas absorption cell as the
velocity metric in October 1990 because of concerns over possible long-term
systematic errors related to the use of the telluric O$_2$  lines.  
Although all of the McDonald data are self-consistent, and the derived orbital
solution for 16~Cygni~B does not depend on the inclusion or exclusion of the
O$_2$ based data, we have decided to restrict the analysis to only the
I$_2$ based data.
The primary reason is that the use of an I$_2$ cell allows modeling of
temporal and spatial variations of the instrumental point-spread function
(\cite{VaBuMa95}).
Such modeling is vital to improve the precision of these measurements.
The McDonald Observations were obtained with the ``6-foot'' camera of the
2.7\,m Harlan J. Smith Telescope coud\'e spectrograph.
All of the Lick data use an I$_2$ cell as the velocity metric.
Lick Observatory data were obtained with the Hamilton Echelle spectrograph, fed
by either the 3\,m Shane Telescope or by the Coud\'e Auxiliary Telescope.

During 1996, both the McDonald and the Lick groups each separately
became convinced of the reality of their observed radial velocity variations
of 16~Cyg~B, and were able to obtain totally independent orbital solutions.
We became aware of each others' work on this star, and found that these
separate orbital solutions agreed to within the uncertainties.
We then decided to combine all of the data into a joint solution.
The measured relative radial velocities from McDonald Observatory
are given in table~\ref{McDvelocities}, and the Lick velocities are given
in table~\ref{Lickvelocities}.
Each of the two separate data sets had an independent and arbitrary zero-point.
In the combined orbital solution, we left the velocity offset between the data
sets as a free parameter.
The values given in Tables~\ref{McDvelocities} and~\ref{Lickvelocities}
have been corrected for this velocity offset. 
Thus, they are on the same zero-point.
The uncertainties for the Lick data are computed from the rms scatter of the
$\sim700$ independent 2{\AA} wide chunks into which the spectrum was divided
for analysis of the spectrograph point-spread function (cf. \cite{BuMaWi96a}).
The McDonald data were obtained at significantly higher spectral resolution ($R
= 210,000$ as opposed to $R = 62,000$ for the Lick data).  However, this
configuration of the McDonald spectrograph was able to record only a 9{\AA}
wide section of a single echelle order near 5200{\AA}.
These McDonald data thus have somewhat lower
measurement precision than the Lick data.
16~Cyg~B is the faintest star on the McDonald program list, and the velocity
precision on this star is limited by photon statistics.
Empirical estimates of the velocity precision obtained from McDonald data as a
function of photon flux agree well with the uncertainties computed following
the derivation of Butler {\it et~al.\/} (1996b).
The error bars on each McDonald measurement listed in table~\ref{McDvelocities}
were computed in this manner from the observed flux in each spectrum.

\section{Orbital Solution}

The weighted orbital solution for the combined Lick and McDonald data is given
in table~\ref{orbit}.  This solution agrees  very well with the orbital
solutions derived separately from each independent data set.
If we adopt a mass for 16~Cyg~B of 1.0M$_{\odot}$ (Friel {\it et~al.\/} 1993),
\nocite{FrCSCh93} this solution gives a planetary orbital semi-major axis
of 1.6 {\sc au}, and $m_P \sin i = 1.5 m_J$.
Figure \ref{rv_curve} shows all of the individual velocity measurements as a
function of time.
The solid line is the radial velocity curve from the orbital solution.
The cyclic repetition of a strongly asymmetric radial velocity variation is
quite obvious from simple inspection of the raw data in figure~\ref{rv_curve}.
This asymmetric, almost sawtooth, variation in radial velocity is a direct
result of the large orbital eccentricity.
The steeply changing portion of the velocity curve corresponds to periastron
passages.
A circular orbit would give a sine wave.

A Lomb-Scargle periodogram analysis (\cite{Sc82}; \cite{HoBa86}) of the
data gives a large peak at a period of 824 days, in excellent agreement
with the orbital solution.
The false-alarm probability of this peak is $2.7 \times 10^{-8}$,
ruling out noise fluctuations as the cause of the observed variations.
Furthermore, it is totally inconceivable that noise would conspire to give
exactly the same apparent false signal in both independent data sets.
The spectral window function shows that this period is not a spectral alias of
some other period.

Figure \ref{16CygA} shows the observed Lick and McDonald radial velocity
variations of 16~Cygni~A, the other star of the wide binary in the 16~Cygni
system.
This star clearly does not show the large radial velocity variations which are
easily evident in the 16~Cygni~B data.
The 16~Cyg~A velocities are consistent with a flat line, indicative of no
variations at all.
We would think that if the observed 16~Cygni~B variations were the result of
some unknown systematic error which somehow managed to give exactly the
same periodic signal in both data sets, then such a systematic error
should also affect observations of 16~Cygni~A, which is only 39 arcsec away
from B and was observed with virtually the same temporal sampling.
A Lomb-Scargle periodogram of the 16~Cygni~A data gives no peak with a
false-alarm probability greater than 0.14, clearly demonstrating the lack of
radial-velocity variability in this star.

The large period and the amplitude of the radial velocity curve of 16~Cygni~B
argue strongly for orbital motion as the cause of the observed velocity
variations.
An integration of the radial velocity curve would imply a a radius
variation of $7.4 \times 10^{10}$\,cm, if one assumes the velocity
variability were due to simple radial pulsations.
This variation is slightly larger than the radius of the star, and is
easily excluded by the lack of any observed photometric variability of
16~Cyg~B.
Moreover, the period of the radial fundamental for a 1.0M$_{\odot}$
main-sequence star is of order 1\,hour (\cite{Cox80}), far different from
the observed 801\,day radial-velocity period.

Non-radial pulsations may be similarly excluded.   Because 16~Cygni~B is a
solar twin,  we would expect it to have a very low-amplitude (less than
1\,m\,s$^{-1}$) p-mode oscillation
spectrum centered near periods of 5\,minutes.   Similarly, its g-mode spectrum
(if present at all) should have periods around 30-40\,minutes,
and amplitudes far less than we observe.
Non-radial pulsations also may be ruled out based on the
lack of spectral line profile variations of an amplitude sufficient to cause
the observed RV variations (cf. \cite{Ha96}; \cite{HaCoJK96}).

The companion object to 16~Cygni~B is quite unlike any other substellar
object found so far.
Figure~\ref{e_vs_m} shows the distribution of orbital eccentricity as a
function of minimum mass for the known substellar companions to solar-type
stars (thus excluding the ``pulsar planets'').
The sample comprises Jupiter and Saturn from our own solar system,
the possible planetary objects discussed in Section~\ref{intro},
as well as brown dwarf companions found in the CfA survey
(\cite{MaLaSt96}) and the Geneva survey (\cite{MaQuUd96}).
This figure immediately shows that the companion to 16~Cygni~B is unique in its
combination of low mass and very large orbital eccentricity.
There is a clustering of objects around Jupiter with low mass and low
eccentricity.
It is tempting to think of all of these as true Jovian planets.
There is a second group of objects with significantly higher masses and a 
wide range of orbital eccentricities -- characteristics that we would expect
for brown dwarf secondaries in binary star systems.
The companion to 16~Cyg~B sits alone in this diagram, with a mass solidly
in the range expected for planets, but with a very large orbital eccentricity.
Its nearest neighbors in this figure, the two ``massive eccentric'' objects
around HD114762 and 70~Vir, have $m \sin i$ five times larger, semi-major axes a
factor of three smaller, and eccentricities about half of the 16~Cyg~B
companion.
This new object around 16~Cyg~B may represent an extreme case of the massive
eccentric planets (perhaps viewed at low $\sin i$), an ultra-low mass brown
dwarf, a ``normal'' planet which was formed in the normal manner with low
orbital eccentricity but was perturbed into a much higher eccentricity,
or it may represent yet another class of planetary-mass companions to
solar-type stars.

\section{Discussion}

The current paradigm for planetary system formation (\cite{PoHuPo93})
which is based on the archetype of our own solar system,
builds planets by accretion processes in a protoplanetary disk surrounding a
newly formed star.   The first step in the formation of Jovian-mass gas-giant
planets is the growth of a rock-ice core in the disk.
The most  massive proto-planetary object will experience a runaway growth,
sweeping up other smaller nearby planetesimals.
When an object reaches a mass of 10-20 earth masses, its gravity is sufficient
to capture gas in the disk, resulting in the very rapid growth of a deep
gaseous envelope.   This general model has been fine-tuned to produce planetary
systems like our own, with a dominant gas-giant planet in a low eccentricity
orbit at about 5~{\sc au}, other smaller gas-giant planets exterior to that,
and smaller rocky bodies interior.

While such a model is probably able to form the pseudo-Jovian planets around
47~UMa and Lalande 21185 with some minor adjustments, this model is unable to
explain either the ``51-Peg'' type planets or the ``massive-eccentric''
systems.  Lin, Bodenheimer, and Richardson (1996) \nocite{LiBoRi96} suggested
that the  51-Peg type planets might be formed by the inward orbital migration
of a gas-giant planet originally formed at much larger distances according to
the conventional paradigm.
This idea has been explored further by Trilling {\it et~al.\/} (1996).
\nocite{TrBeGu96}  
Tidal interactions will transfer angular momentum
from the planet to the disk exterior to its orbit, causing it to spiral slowly
toward the star.  The inward migration is stopped at about 0.05~{\sc au} either
by tidal interactions with the spin of the star or by the clearing of the inner
disk by the stellar magnetosphere.

Both of these mechanisms will produce planets in low eccentricity orbits.
Indeed, any planet formed in a classical circumstellar disk should start its
life in a nearly circular orbit.
Tidal interactions between the disk and the planet will tend to circularize the
orbit quickly.
Stellar companion objects, however, can have a very wide range of
eccentricities and semi-major axes (\cite{DuMa91}).
One possible explanation of these three low-mass eccentric objects is that they
simply represent the low mass end of the brown-dwarf mass function.
If this is true, then this mass function is remarkably flat through the range
0.001--0.080\,M$_{\odot}$, and the process of binary star formation can produce
systems with mass rations of $10^2$ to $10^3$.

An alternative explanation for the formation of the massive eccentric planets
was suggested by Lin and Ida (1996). \nocite{LiId96}
They demonstrated that if several Jovian-mass objects can form within the
context of the classical paradigm at semimajor axes greater than
$\sim$1\,{\sc au} in a somewhat massive disk, then such a system might evolve
into the types of systems we find in 70~Vir and HD114762.  
Numerical integrations of the orbital evolution showed that
such systems will be stable while the disk is still present, but after disk
dissipation the mutual gravitational perturbations of these
planets on each other will often cause the system to become chaotic.
The orbital eccentricities of the planets will increase and
their orbits will begin to cross.  At that point, the system will evolve
rapidly, as the planets will collide and merge.
The end result is typically a massive inner planet with a large eccentricity
and small semi-major axis, sometimes accompanied by other planets in
much more distant orbits.
Such a scenario might provide a plausible explanation for the low-mass
eccentric companion we have found to 16~Cyg~B.
Alternatively, large eccentricities in massive planets 
may be excited by the protoplanetary disk itself.  Lindblad 
resonances in the disk material may overcome the usual dynamical damping
to augment the eccentricity (\cite{ArLu96}).

The presence of the companion to 16~Cyg~B around the secondary star in a binary
system offers interesting additional possible explanations for its large
eccentricity.   The current projected separation of 16~Cyg~A and B is about
39\,arcsec, which corresponds to 835\,{\sc au} at the parallax of
0.0467\,arcsec (\cite{vALeHo91}).   The orbit of 16~Cyg~A and B about each
other is very poorly determined, as reliable astrometric data covers only a
very short arc of the orbit.
An orbital solution was attempted by Romanenko (1994) \nocite{Ro94} using the
method of apparent-motion parameters, \nocite{Ro94} but it is uncertain
how reliable this solution is.   
Most of the analytic investigations into the stability of planets in binary
star systems have been within the context of the restricted three-body problem.
Many of the later numerical studies have considered rather short integrations.
According to the classical analytical studies, the planet around 16~Cyg~B
should be stable as long as the stellar semi-major axis is greater than about
10 {\sc au}, which is certainly the case given the present separation of the
stars.  A more detailed numerical investigation by Holman and Wiegert (1996)
\nocite{HoWi96} derived an expression for the critical planetary semimajor
axis for stability as a function of the binary eccentricity, mass ratio, and
semi-major axis.  Applying this to the 16 Cyg system, the planet should be
stable (i.e. it does not become unbound) for almost all plausible values of
stellar semi-major axis and eccentricity, provided the relative inclination
of the stellar and planetary orbits is not extremely large.  
However, as was pointed
out by Wiegert and Holman (1996) \nocite{WiHo96} in their study of the
stability of planets in the $\alpha$~Centauri system, if the inclination of the
planetary orbit with respect to the stellar orbit is near $90^\circ$, then the
planet can easily be lost from the system.
Even if the planetary orbit is ``stable'', it is still quite possible for the
stellar companion to strongly influence the evolution of the planetary orbit.
In cases where there is an inclination between then stellar and planetary
orbital planes, the planetary orbit will suffer an exchange of energy between
inclination and eccentricity, with the semi-major axis
remaining approximately constant, an effect first discussed by Kozai (1962).
\nocite{Ko62}
This mechanism has been explored in detail by Holman {\it et~al.\/} (1996)
and by Mazeh {\it et~al.\/} (1996). \nocite{HoToTr96,MaKrRo96}
Whether this effect is responsible for the large eccentricity of the planet
around 16~Cyg~B is unknown because of the large uncertainties about
the parameters of the stellar orbit and the relative inclinations of the
stellar and planetary orbital planes.
Hale (1994) \nocite{Ha94} examined the question of coplanarity between the
orbital and the stellar equatorial planes in solar-type binary systems.
This study concluded that systems with large ($\> 100$ {\sc au}) separations
showed little correlation between the orbital plane and the stellar equatorial
planes, while systems with separations less than 30-40 {\sc au} showed
approximate coplanarity.
For the specific case of 16~Cygni~A and B, Hale estimated stellar equatorial
inclinations of ${43^{\circ}}^{+47}_{-29}$ and ${90^{\circ}}^{+0}_{-34}$
respectively.
While the orbital inclination of the binary is unknown,
from the statistical results of
Hale's study it is unlikely that it coincides with the rotational inclination
of either star, and thus probably does not coincide with the original
orbital inclination of the planetary companion to 16~Cyg~B.
Thus, this mechanism provides a quite plausible explanation for
the large observed eccentricity of this object.
Any additional observational constraints on these parameters would be of
immense value.

An alternate possible scenario for the origin of the large planetary
eccentricity is also tied to the formation of the planet in a multiple star
system.   Such systems often form as ``trapezium'' systems of several stars in
a dynamically unstable configuration.
A three-body gravitational interaction often will eject one object from the
system, and will often bind the other two stars in a tighter binary orbit.
The 16~Cygni system is now a triple star system,
with a distant M~dwarf companion to the G~dwarf binary.   It is quite possible
that the system originally had some other component when it was formed.
A three-body encounter between A, B, and this other, now ejected, star could
have served to perturb the planet around B into its present large eccentricity
orbit.   Unfortunately, such a scenario is virtually impossible to prove.

Analysis of the photospheric lithium abundance may provide interesting
additional constraints on the formation and early evolution of this system.
Lithium is easily destroyed by (p,$\alpha$) nuclear processes at
relatively low temperatures of a few million degrees, which can be reached at
the bottom of the convective zone in G~dwarfs.
The lithium abundance of a solar-type dwarf
can provide limits on the depth and extent of convective mixing of photospheric
material in the formation and early (pre-main sequence) evolution of the star.
Lithium in the sun is depleted by a factor of 100 relative to the meteoritic
abundance.
Observations by King {\it et~al.\/} (1996) \nocite{KiDeHi96} show that 16~Cyg~A
has a mean photospheric Li abundance of $\log N(\rm{Li}) = 1.27 \pm 0.04$,
while the abundance of 16~Cyg~B is $\log N(\rm{Li}) = 0.48 \pm 0.14$.
The solar value, computed with the same analysis procedure, is intermediate
between these two stars at $\log N(\rm{Li}) = 1.05 \pm 0.06$.
We confirm, from the Lick spectra, that the Li 6707{\AA} resonance
line in 16~Cyg~A is considerably
stronger ($W_{\lambda} = 12$m{\AA}) than that in B.
As was discussed in Section~\ref{obs}, the spectra of the Sun, 16~Cyg~A, and
16~Cyg~B are otherwise nearly indistinguishable.
This difference in the Li abundance between 16~Cyg A and B most likely points
to a difference in the mixing of the photospheres and the deeper layers of
these stars, which is probably driven by rotation (\cite{PiKaDe90}).
Thus, it is likely that 16~Cyg~A and B had a somewhat different angular
momentum history, either a difference in the initial angular momentum or a
difference in the rate at which the stellar rotation has slowed.

Stars of equal mass, age, and initial chemical composition may nonetheless
differ in angular momentum.   The angular momentum history of  young solar-type
stars is governed strongly by torques exerted on the star by the inner
accretion disk, as is observed in pre-main-sequence (PMS) stars
(\cite{EdStHa93}).  PMS stars with massive disks exhibit slow rotation rates due
to the (presumably magnetic) coupling to the inner disk (\cite{ArCl96}).
Observationally, slowly rotating young G stars exhibit lower photospheric Li
abundances, and hence have burned Li more rapidly than the rapid rotators
(\cite{SoJoBa93}).   This effect is opposite to that predicted by early models
of Li depletion, involving meridonal circulation.
More recent work by Mart\'in and Claret (1996) \nocite{MaCl96} shows that rapid
rotation in contracting PMS stars can significantly inhibit the depletion of
Li, in accord with observations.   Thus, we now have a reasonably coherent
connection between the presence of a disk, early stellar rotation rates, and
Li depletion.    The large spread in photospheric Li abundance at a given mass
and age in well-studied young clusters is a result of the spread in rotation
rates of these stars during their PMS phase.   The rotation rates are, in turn,
regulated by the disk mass (\cite{St94}).   Therefore, it may be possible to
use the Li abundance of an older main-sequence star as a diagnostic of the
mass of its protoplanetary disk, and hence of the planet-building environment.
From this picture, we would expect that solar-type stars with large Li
abundances for their mass and age were rapidly rotating in their PMS stage due
to the lack of rotational braking by a massive proto-planetary disk.
Conversely, a star being depleted in Li for its mass and age  would indicate
slow PMS rotation resulting from the presence of a proto-planetary disk.
This reasoning suggests that for stars of equal mass and age, one may rank
their proto-planetary disk masses based on the current-epoch Li abundances.
The sense of this is consistent with what we see in 16 Cyg A and B.
We find significant Li depletion in 16~Cyg~B, and this star has a 1.5 Jupiter
mass planet at 1.6\,{\sc au} which presumably formed in a massive
protoplanetary disk which would have served to brake the stellar rotation.
On the other hand, 16~Cyg~A has a much larger Li abundance, indicating more
rapid PMS rotation.
We find no Jovian mass companion to 16~Cyg~A, which is consistent with the lack
of a massive protoplanetary disk to provide rotational braking for the star.
If the Sun has an age similar to that of the 16~Cyg system, we would conclude
that it had a disk intermediate between those of 16 Cyg A and B.
Indeed, the major planet around the Sun is of lower mass and is farther
from the star than in the case of 16~Cyg~B.

\acknowledgements
Work at McDonald Observatory was supported by NASA grant NAGW-3990.
The Lick Observatory program was supported by NASA grant NAGW-3182, by
NSF grant AST95-20443, and by Sun Microsystems.
We would like to thank Jeremy King and Jihad Touma for enlightening discussions.
WDC is very grateful to the late Harlan J. Smith for his enthusiastic support
and encouragement of the McDonald program during its early years, and for his
faith that radial velocity techniques would find an abundance of planetary
systems around nearby stars.
We all wish that he could have lived to see this vision come true.

\clearpage
\begin{figure}
\caption{The combined Lick and McDonald radial velocities for 16 Cygni B.
The triangles are from Lick data and the crosses are from McDonald.
The solid line is the radial velocity curve from the orbital solution.}
\label{rv_curve}
\end{figure}

\begin{figure}
\caption{The combined Lick and McDonald radial velocities for 16 Cygni A.
The symbols are the same as in Fig.~\ref{rv_curve}.
No radial velocity variation is evident in these data.}
\label{16CygA}
\end{figure}

\begin{figure}
\caption{The relationship between eccentricity and minimum mass for the 
substellar companions (both planets and brown dwarfs) found so far.}
\label{e_vs_m}
\end{figure}

\clearpage

\begin{deluxetable}{lcccl}
\tablecaption{Comparison of Physical Parameters of the Sun with 16~Cygni~A and B
\label{star_properties}}
\tablewidth{0pt}
\tablehead{
\colhead{Parameter} & \colhead{Sun} & \colhead{16 Cygni A} &
\colhead{16 Cygni B} & \colhead{Reference}
}
\startdata
Spectral Type         & G2V     & G1.5V        &G2.5V & \\
$T_{eff}$ (K)         & 5770    & 5785$\pm$25  &5760$\pm$20   &\cite{FrCSCh93}\\
$\log g$ (cgs)        & 4.44    & 4.28$\pm$0.07&4.35$\pm$0.07 &\cite{FrCSCh93}\\
Mass (M$_{\odot}$)    & 1.0     & 1.05$\pm$0.05&1.00$\pm$0.05 &\cite{FrCSCh93}\\
$[{\rm Fe}/{\rm H}]$  & 0.0     &+0.05$\pm$0.06&+0.05$\pm$0.06&\cite{FrCSCh93}\\
$v\sin i$(km s$^{-1}$)&1.9$\pm$0.3&1.6$\pm$1.0 &2.7 $\pm$ 1.0 &\cite{So82}\\
Rotation Period (days)& 25.38   & 26.9         & 29.1         &\cite{Ha94} \\
\enddata
\end{deluxetable}

\clearpage
\begin{deluxetable}{rrrrrr}
\tablecaption{McDonald Relative Radial-Velocities of 16~Cygni~B
\label{McDvelocities}}
\tablewidth{0pt}
\tablehead{
\colhead{JD-2400000.0}&\colhead{V (m\,s$^{-1}$)}&\colhead{$\sigma$(m\,s$^{-1}$)}&
\colhead{JD-2400000.0}&\colhead{V (m\,s$^{-1}$)}&\colhead{$\sigma$ (m\,s$^{-1}$)}
}
\startdata
48485.7305 &   -4.2 &  16.8 &
48524.6562 &   22.6 &  21.9 \\
48783.8398 &   41.5 &  20.5 &
48823.8984 &    8.7 &  16.7 \\
48852.7734 &   22.2 &  21.3 &
48882.5742 &   52.7 &  19.6 \\
48901.6875 &    5.8 &  20.4 &
48943.6328 &  -13.2 &  22.8 \\
49146.9102 &  -37.3 &  22.1 &
49220.7812 &  -10.8 &  19.5 \\
49258.7500 &   -5.1 &  20.5 &
49286.6367 &  -27.9 &  17.1 \\
49521.8984 &   20.5 &  21.4 &
49588.7305 &   51.5 &  16.0 \\
49616.6602 &   45.3 &  24.2 &
49647.6289 &   44.4 &  14.6 \\
49668.5430 &   13.7 &  22.6 &
49703.5508 &   35.7 &  18.1 \\
49816.9023 &  -37.6 &  21.0 &
49861.9141 &  -13.6 &  19.5 \\
49876.8438 &   -6.9 &  21.9 &
49916.7930 &   -7.6 &  17.2 \\
49946.8125 &  -39.6 &  19.4 &
49963.7070 &  -33.6 &  16.1 \\
49994.6055 &   -5.0 &  16.4 &
50204.9258 &   19.7 &  21.6 \\
50235.9023 &   38.5 &  19.6 &
50292.7539 &   20.9 &  22.8 \\
50355.6328 &   45.5 &  19.0 \\
\enddata
\end{deluxetable}

\clearpage
\begin{deluxetable}{rrrrrr}
\tablecaption{Lick Relative Radial-Velocities of 16~Cygni~B
\label{Lickvelocities}}
\tablewidth{0pt}
\tablehead{
\colhead{JD-2400000.0}&\colhead{V (m\,s$^{-1}$)}&\colhead{$\sigma$(m\,s$^{-1}$)}&
\colhead{JD-2400000.0}&\colhead{V (m\,s$^{-1}$)}&\colhead{$\sigma$ (m\,s$^{-1}$)}
}
\startdata
47046.7695 &   17.9 &   6.8 &
47846.6758 &   41.9 &  10.7 \\
48019.9531 &   55.4 &  12.0 &
48113.8242 &   61.1 &  10.7 \\
48438.8750 &  -10.0 &   8.9 &
48846.8750 &   45.6 &   9.1 \\
48906.6992 &   56.5 &   9.2 &
49124.9844 &  -38.9 &   8.2 \\
49172.9062 &   -2.6 &   4.6 &
49200.8750 &  -22.3 &   9.8 \\
49588.7422 &   28.5 &   7.7 &
49601.7656 &   17.8 &   8.1 \\
49623.7266 &   47.5 &   9.6 &
49858.9844 &  -37.1 &   6.4 \\
49892.9648 &  -20.4 &   5.5 &
49914.9688 &  -17.8 &   4.6 \\
50069.5781 &  -20.4 &   8.7 &
50072.5742 &  -11.9 &   5.1 \\
50073.5977 &  -11.7 &   4.0 &
50089.5938 &   -6.8 &   5.9 \\
50182.0039 &    1.2 &   4.3 &
50202.0039 &    9.7 &   8.8 \\
50203.9648 &   -3.1 &   6.2 &
50215.9336 &    9.8 &   5.8 \\
50231.9219 &   -9.0 &  12.9 &
50235.9883 &   -0.8 &   8.7 \\
50262.9023 &   10.4 &   4.7 &
50288.7578 &   12.1 &   4.7 \\
50298.7852 &   17.9 &   3.5 &
50299.9414 &   20.8 &   3.2 \\
50300.7305 &   20.1 &   2.8 &
50300.9375 &   16.9 &   3.7 \\
50304.6758 &   23.9 &   2.8 &
50304.9492 &   21.4 &   2.9 \\
50305.7539 &   18.5 &   4.0 &
50307.7969 &   21.8 &   4.0 \\
50309.8008 &   21.6 &   3.6 &
50311.7695 &   12.6 &   3.6 \\
50326.7734 &   13.7 &   3.0 &
50372.6836 &   25.7 &   4.2 \\
50377.7031 &   28.1 &   4.4 &
\enddata
\end{deluxetable}
 
\clearpage
\begin{deluxetable}{lcc}
\tablecaption{Combined weighted Orbital Solution for 16~Cygni~B
\label{orbit}}
\tablewidth{0pt}
\tablehead{
\colhead{Parameter} & \colhead{Value} & \colhead{Uncertainty}
}
\startdata
Orbital Period $P$ (days) &  800.8 & 11.7 \\
Velocity Semi-amplitude $K$ (m\,s$^{-1}$) & 43.9  & 6.9 \\
Eccentricity $e$ & 0.634 & 0.082 \\
Longitude of Periastron $\omega$ (degrees) & 83.2 & 12.7 \\
Periastron Date $T_0$ (Julian Date) & 2448935.3 & 12.0 \\
\enddata
\end{deluxetable}
 
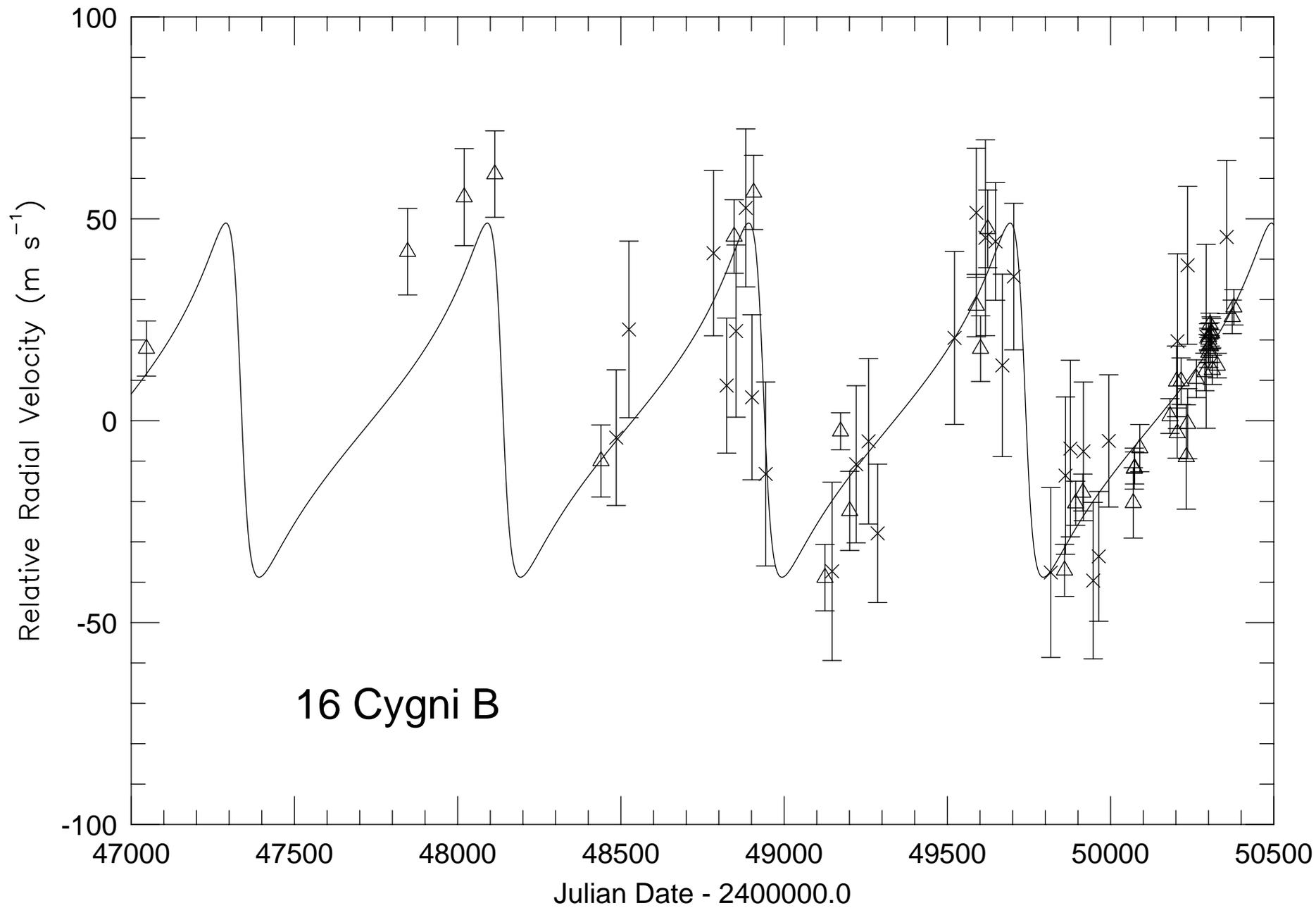
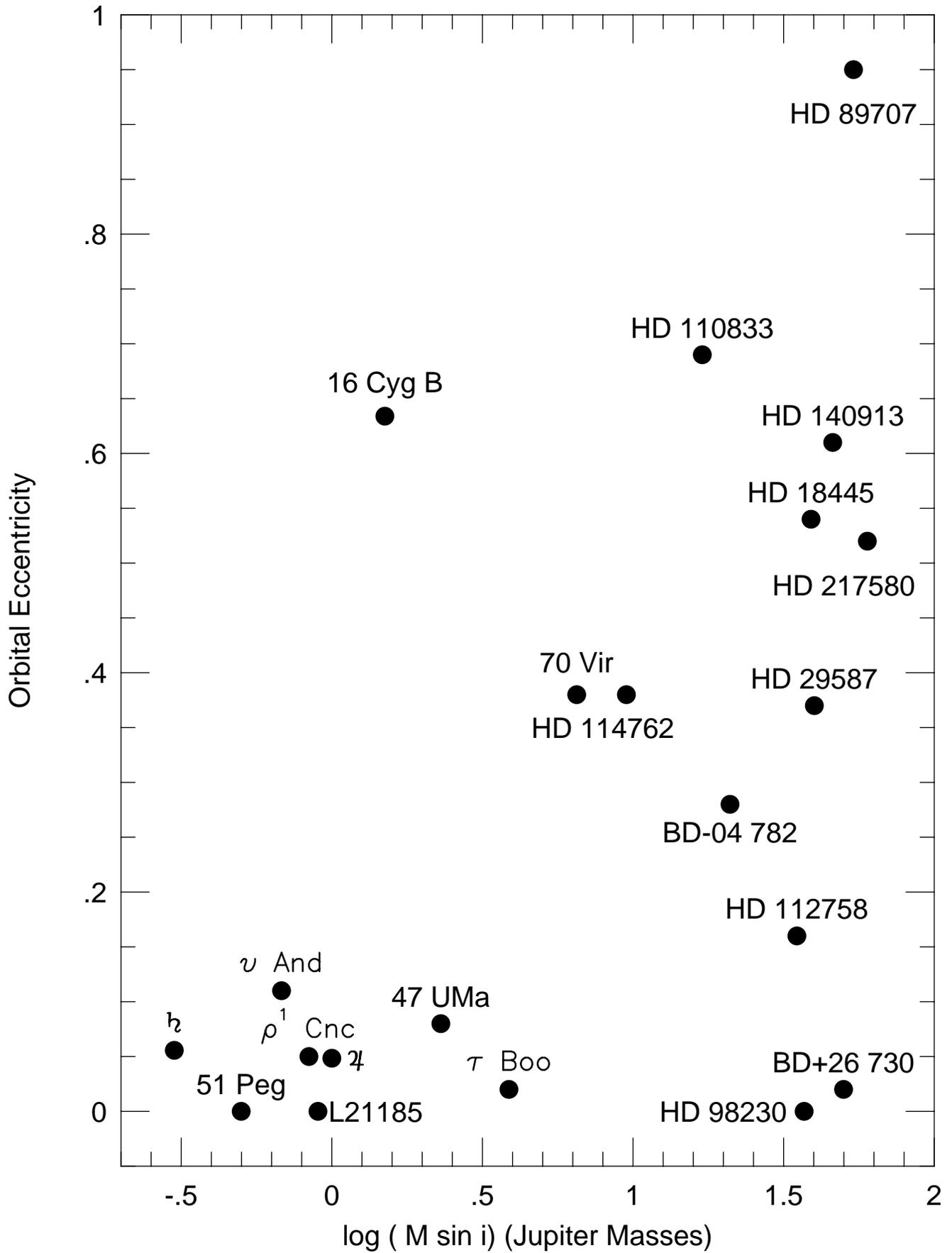
\end{document}